\documentclass{article}
\usepackage{spconf,amsmath,epsfig}
\usepackage{cite}
\usepackage{amsmath,amssymb,amsfonts}
\usepackage{algorithmic}
\usepackage{graphicx}
\usepackage{textcomp}
\usepackage{xcolor}
\usepackage{booktabs}
\usepackage{float}
\usepackage{fancyhdr}

\title{3D Super-Resolution Imaging Method for Distributed Millimeter-wave Automotive Radar System}
\name{Yanqin Xu, Xiaoling Zhang, Shunjun Wei, Jun Shi, Xu Zhan, Tianwen Zhang}
\address{School of Information and Communication Engineering\\
	University of Electronic Science and Technology of China
	\\Chengdu, China, 611731\\Email:202111012112@std.uestc.edu.cn
}
\begin{document}
%
\maketitle
\begin{abstract}
 
Millimeter-wave (mmW) radar is widely applied to advanced autopilot assistance systems. However, its small antenna aperture causes a low imaging resolution. In this paper, a new distributed mmW radar system is designed to solve this problem. It forms a large sparse virtual planar array to enlarge the aperture, using multiple-input and multiple-output (MIMO) processing. However, in this system, traditional imaging methods cannot apply to the sparse array. Therefore, we also propose a 3D super-resolution imaging method specifically for this system in this paper. The proposed method consists of three steps: (1) using range FFT to get range imaging, (2) using 2D adaptive diagonal loading iterative adaptive approach (ADL-IAA) to acquire 2D super-resolution imaging, which can satisfy this sparsity under single-measurement, (3) using constant false alarm (CFAR) processing to gain final 3D super-resolution imaging. The simulation results show the proposed method can significantly improve imaging resolution under the sparse array and single-measurement.
\end{abstract}
\section{Introduction}
\label{sec:intro}
As a microwave sensor, the mmW radar has the advantages of small size, low price, and strong environmental adaptability all day long, which has been widely used in autopilot assistance systems [1]. It can form a 3D imaging (range-azimuth-elevation) to perceive the surrounding environment [2]. It has already achieved centimeter-level range resolution (3.75 cm with 4 GHz sweep bandwidth) by using 77 GHz frequency modulated continuous wave (FMCW) [3]. Nevertheless, the most challenging problem is how to achieve a higher azimuth-elevation resolution.

In order to solve this problem, much research has been carried out. The most straightforward way is increasing the antenna aperture. For instance, the MIMO radar system can make the $N$ transmitting antennas and $M$ receiving antennas form a virtual aperture of $NM$ array antennas elements through time division multiplexing(TDM) to increase antenna aperture [4]. The synthetic aperture radar (SAR) technology can form a larger aperture by the precise antenna movement. SAR technology usually needs to be compensated for movement errors [5]. Due to the short wavelength, mmW automotive radar needs some complex algorithms to accurately compensate for this error, moreover, the automotive radar can only get side-looking imaging by SAR technology. Therefore, we designed a new distributed mmW radar system, which can effectively enlarge the aperture.

On the other hand, various super-resolution techniques algorithms have been widely used in the imaging field, like multiple signal classification(MUSIC) and Capon beamformer [6], and so on can effectively improve the azimuth and elevation resolution. Regrettably, these algorithms generally need a large number of measurements. But in fact, the number of snapshots is limited in mmW automotive radar real-time processing under only a single-measurement, which the performance of these algorithms is observably reduced. Meanwhile, these algorithms cannot apply to the sparse antenna aperture. Compressive sensing (CS) is able to efficiently reconstruct imaging in sparse arrays with the same specific parameters [7], but in real-time processing, these specific parameters are difficult to meet. Fortunately, a nonparametric iterative adaptive approach can be used in an arbitrary array with low side lobes [8]. It inspired us to discuss applying a 2D ADL-IAA algorithm to satisfy the sparsity and single-measurement condition in this system.

In this paper, a new distributed mmW radar system is designed first. Then, we propose a 3D super-resolution imaging method for this system: (1) the range FFT to make the echoes of the sparse array to obtain range imaging firstly. (2) Next, the proposed 2D ADL-IAA algorithm to get azimuth-elevation super-resolution imaging on each range unit of range imaging. (3) Finally, an appropriate threshold for CFAR is used to get the last 3D super-resolution imaging, included range-azimuth-elevation resolution. 
 \begin{figure}[H]
 	\centerline{\includegraphics[height=3.5cm]{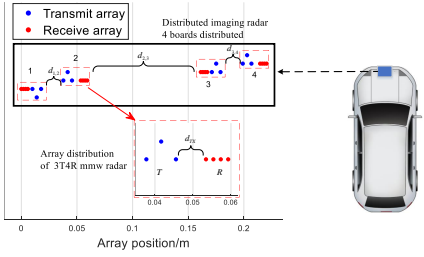}}
 	\caption{Distibuted radar sensor model.}
 	\label{fig1}
 \end{figure}
 \begin{figure}[H]
	\centerline{\includegraphics[height=3.5cm]{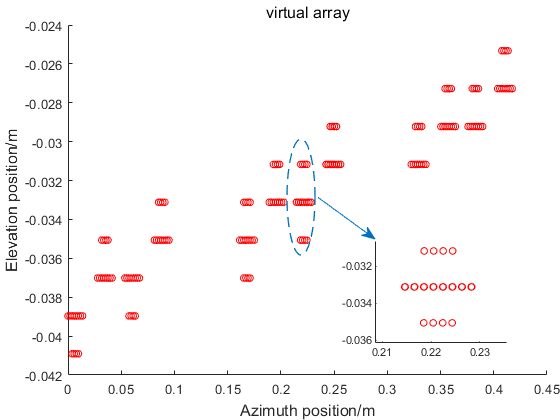}}
	\caption{Vritual planar array.}
	\label{fig2}
\end{figure}
\section{Distributed  automotive radar system}
\label{sec:format}
\subsection{Design Principle}
\label{ssec:subhead}
We main reference to the current 3 transmitting 4 receiving (3T4R) mmW radar common size: the distance of receive and transmit array is $d={\lambda }/{2}\;$ and the distance between the transmit and receive array is ${{d}_{TX}\approx0.41}$cm. The distributed radar system is designed by using four of the same radars in Fig.~\ref{fig1}. After TDM-MIMO processing, the sparse 2D virtual planar antenna array is shown in Fig.~\ref{fig2}.

 We consider three design conditions: \textit{1)} minimize sparsity, \textit{2)} azimuth resolution capability, \textit{3)} radar size. So the design principle is as follows: first, we choose 1, 2 radar to be a group and 3, 4 radar to be a group. Next, the radars between the groups are arranged face-to-face and outside the group are arranged back-to-back. Therefore, the distance between each radar is: ${{d}_{1,2}\approx2}$cm, ${{d}_{2,3}\approx10.1}$cm, ${{d}_{3,4}\approx2}$cm. Then, in order to endow elevation resolution, we stagger each radar at ${\lambda }/{2}\;$. Note that, the distributed radar system can be synchronous by connecting cables and using a central control system to ensure proper coherent processing across multiple radars.

\subsection{ Reslution Analysis}

\begin{figure}[H]
	\centering
	\begin{minipage}[!]{0.4\linewidth }
		\centering
		\includegraphics[width= \linewidth,height=0.12\textheight]{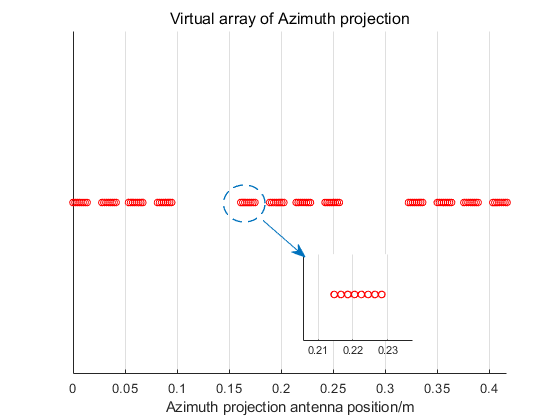}
		\begin{footnotesize}
			(a)
		\end{footnotesize}   
	\end{minipage}%
	\begin{minipage}[!]{0.4\linewidth }
		\centering
		\includegraphics[width= \linewidth,height=0.12\textheight]{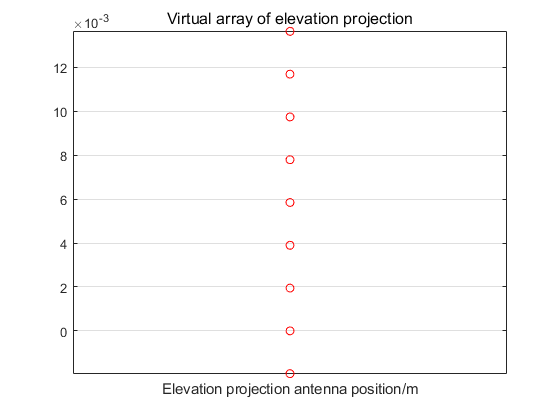}
		\begin{footnotesize}
			(b)
		\end{footnotesize}   
	\end{minipage}%
	\caption{Virtual projection array of azimuth and elevation: (a) azimuth projection array, (b) elevation projection array.}
	\label{fig3}
\end{figure}

Next, we carried out azimuth-elevation resolution analysis of the azimuth-elevation projection of the virtual array in Fig.~\ref{fig3}.
\textit{1) Azimuth resolution:} The number of the smallest redundant azimuth array elements is $M_a=128$, so we can calculate the length of aperture $D\approx 42$cm in Fig.~\ref{fig3}(a).  According the  resolution equation [9], the azimuth resolution can be achieved at $\Delta \theta \approx {{0.5}^{\circ }}$.

\textit{2) elevation resolution:} Since the elevation array elements are uniform and elements $M_e=9$, so, $D=M_e\lambda/2$ in Fig.~\ref{fig3}(b). The elevation resolution is ${{12}^{\circ }}$.

\section{3D Super-Resolution  Imaging Method}

\subsection{Signal Model}
The linear FMCW signal has been adapted in mmW radar effectively. The FMCW signal is expressed as
\begin{equation}
	\ s_{tr}\left( t \right)=\exp \left( j2\pi \left( {{f}_{0}}t+ 0.5K{{t}^{2}} \right) \right),0\le t\le T\label{eq2}
\end{equation}
where $f_0$ is chirp frequency, $K=B/T$ is chirp slope, $B$ and $T$ respectively are sweep bandwidth and pulse duration.

The receiving  antennas the FMCW signal that is reflected in targets, and after mixing processing, the echo is given as
\begin{equation}
	\ {{s}_{re}}\left( {{t}_{n}} \right)=\exp \left( -j2\pi K\tau {{t}_{n}} \right)\exp \left( -j2\pi {{f}_{0}}\tau  \right),n=1,\cdots ,N\label{eq4}
\end{equation}
where $\tau=2R/c$ is a delay of  target-to-radar and $R$ is a distance of target-to-radar, $t_n$ is fast samples time, and $N$ is samples amount at a fast time.

Performing range FFT processing on \eqref{eq4} to get distance information:
\begin{equation}
	\begin{aligned} 
		{{F}_{ra}}&=FI\left[ {{s}_{re}}\left( {{t}_{n}} \right) \right] \\
		& \approx \sin c\left( R \right)\cdot \exp \left( -j2\pi {{f}_{0}}\tau  \right)  \label{eq5}
	\end{aligned}
\end{equation}
where $FI[\cdot ]$ represents FFT operation, $\sin c(R)$ represents the sampling function  at the peak value at $R$. 
\subsection{2D ADL-IAA Principle}

As shown in Fig.~\ref{fig2}, the 2D virtual planar array is made up of  $M$ virtual antenna elements. Considering that the signals incident into the planar array in $K$ directions from the far-field. The 2D planar array receiving signals is given:
\begin{equation}                                                                 
	\ \boldsymbol{x}_{a}\left( n \right)=\sum\limits_{k=1}^{K}{\boldsymbol{a} ( {{\theta }_{k}},{{\varphi }_{k}} )}{{s}_{re}}\left( n \right)+\boldsymbol{v}\left( n \right)\label{eq6}
\end{equation}
where $\theta_k$ and $\varphi_k$ are azimuth-elevation of the $kth$ signal, $\boldsymbol{a}\left( {{\theta }_{k}},{{\varphi }_{k}} \right)={{\left[ 1,{{e}^{-j{{\phi }_{1}}}},\cdots ,{{e}^{-j{{\phi }_{M-1}}}} \right]}^{T}}$ is  steering vector of the $kth$ signal. The spatial phase difference is ${{\phi }_{m}}\left( {{\theta }_{k}},{{\varphi }_{k}} \right)=2\pi \left( {{x}_{m}}\sin {{\theta }_{k}}\cos {{\varphi }_{k}}+{{y}_{m}}\sin {{\theta }_{k}}\sin {{\varphi }_{k}} \right)/\lambda \ $, ${{d}_{m}}\left( {{x}_{m}},{{y}_{m}} \right)\in \left( 0,M \right)$ is the $mth$ antenna element location, $\lambda=c/f_0$ is the wavelength, $\boldsymbol{v}(n)$ is the Gaussian noise vector of $M$ antenna elements, and the mean is zero, the variance is ${{\sigma }^{2}}$.And $n$ represents the $nth$ sample of fast time.

Due to the stationarity of the received signal, we can get the spatial covariance matrix with $N$ samples of fast time.
\begin{equation}
	\begin{aligned} 
		\boldsymbol{R}&=\boldsymbol{A}\boldsymbol{P}{{\boldsymbol{A}}^{H}}+\sigma^2 \boldsymbol{I} \\ 
		&\approx \frac{1}{N}\sum\limits_{n=1}^{N}{\boldsymbol{x}\left( n \right)}{\boldsymbol{x}^{H}}\left( n \right) \label{eq8}
	\end{aligned}
\end{equation}
where $\boldsymbol{P}=diag\{{{P}_{1}},\cdots ,{{P}_{K}}\}\in {\mathbb{C}}^{K\times K}$ is diagonal matrix of signal power, and ${{P}_{k}}=\frac{1}{N}\sum\limits_{n=1}^{N}{{{\left| {{s}_{k}}\left( n \right) \right|}^{2}}}$ is the $kth$ signal average power.

And a weighted least squares approach for the $kth$ direction:
\begin{equation}
	\min \sum\limits_{n=1}^{N}{\left\| \boldsymbol{x}(n)-{{s}_{k}}(n)\boldsymbol{{a}}({{\theta }_{k}},{\varphi }_{k}) \right\|_{{{\boldsymbol{B}}^{-1}}}^{2}}\label{eq9}
\end{equation}
where $\left\| \boldsymbol{x} \right\|_{{{\boldsymbol{B}}^{-1}}}^{2}\triangleq {{\boldsymbol{x}}^{H}}{{\boldsymbol{B}}^{-1}}\boldsymbol{x}$, $\boldsymbol{x}$ is recieving signal and ${s_k(n)}$ is the $kth$ direction, and definition the equation: $\boldsymbol{B}=\boldsymbol{R}-P_k\boldsymbol{{a}}({{\theta }_{k}},{\varphi }_{k}){\boldsymbol{{a}}^{H}({{\theta }_{k}},{\varphi }_{k})}$.

Then, minimizing \eqref{eq9} with respect to $s_k(n)$ is given:
\begin{equation}
	\ {{s}_{k}}\left( n \right)=\frac{{{a}^{H}}\left( {{\theta }_{k},{\varphi }_{k}} \right){{R}^{-1}}\boldsymbol{x}\left( n \right)}{{{a}^{H}}\left( {{\theta }_{k}},{\varphi }_{k} \right){{R}^{-1}}{{a}^{H}}\left( {{\theta }_{k}},{\varphi }_{k} \right)}\label{eq10}
\end{equation}

The algorithm solution consists of defining a scanning grid of $L$ directions by constructing a measurement matrix
$\boldsymbol{A}\left( \theta ,\varphi  \right)=\left[ \boldsymbol{a}\left( {{\theta }_{1}},{{\varphi }_{1}} \right),\cdots ,\boldsymbol{a}\left( {{\theta }_{L}},{{\varphi }_{L}} \right) \right]$. Due to the 2D estimation needed to cover the whole space, the large number of azimuth-elevation scanning grids is $L$ that leads to $L$-dimensional $\boldsymbol{R}$ matrix inverse problems. Considering the inverse problem of matrix $\boldsymbol{R}$ under large dimensions, we ues the adaptive diagonal loading  algorithm to solve the problem [10]:
\begin{equation}
\ {q} = \frac{1}{L}\sum\limits_{l=1}^{L}{\left| \frac{\boldsymbol{I}_{l}^{H}{\boldsymbol{R}^{\left( i-1 \right)}}^{\text{-}1}{{x}_{a}}\left( n \right)}{\boldsymbol{I}_{l}^{H}{\boldsymbol{R}^{\left( i-1 \right)}}^{\text{-}1}{\boldsymbol{I}_{l}}} \right|} ^{2}\label{eq99}
\end{equation}
By rearranging $\boldsymbol{P}$, we can get azimuth-elevation 2D imaging, and the algorithm can work well in Fig.~\ref{fig2}. The 2D ADL-IAA is summarized in algorithm 1.

\subsection{An Appropriate Threshold for CFAR Processing}
 In the previous section, we can get a 3D matrix by applying 2D ADL-IAA in the $N$ point range unit. Because the proposed algorithm has lower side lobes, we can easily use an appropriate threshold for sensitive CFAR processing to detect targets on background noise and clutter. In the following simulations, we use a probability of false alarm  ${10}^{-4}$ for the CFAR detector.
 \begin{table}[H]
 	\centering
 	\begin{tabular}{llll}
 		\toprule
 		&\textbf{Algorithm 1: 2D ADL-IAA} \\
 		\midrule
 		&\textbf{Input} recieved signal $\boldsymbol{x}_a(n)$, iteration accuracy $\varepsilon$\\
 		&Initialize ${{{P}}_{l}}=\frac{{{\left| {{\mathbf{a}}^{H}}\left( {{\theta }_{l}},{{\varphi }_{l}} \right)\boldsymbol{x}_a(n) \right|}^{2}}}{{{\left( {{\mathbf{a}}^{H}}\left( {{\theta }_{l}},{{\varphi }_{l}} \right)\mathbf{a}\left( {{\theta }_{l}},{{\varphi }_{l}} \right) \right)}^{2}}}$, iteration $i = 0$ \\
 		&1$\,$ \textbf{for} $i$, ${{\boldsymbol{{R}}}^{(0)}}=\boldsymbol{A}\left( \theta ,\varphi  \right)\boldsymbol{P}{{\boldsymbol{A}}^{H}}\left( \theta ,\varphi  \right)+{{\sigma }^{2}}\boldsymbol{I}$\\
 		&2$\quad$ \textbf{for} $l=1,2,\cdots ,L$\\
 		&3$\quad$ $\quad$ ${{{{s}}}_{l}}(n)=\frac{{{\boldsymbol{a}}^{H}}\left( {{\theta }_{l}},{{\varphi }_{l}} \right){{\boldsymbol{R}}^{\left( i\right){\text{-}1}}}\boldsymbol{x}_a(n)}{{{\boldsymbol{a}}^{H}}\left( {{\theta }_{l}},{{\varphi }_{l}} \right){{\boldsymbol{R}}^{\left( i\right){\text{-}1}}}\boldsymbol{a}\left( {{\theta }_{l}},{{\varphi }_{l}} \right)}$  \\
 		&4$\quad$ \textbf {end}\\ 
 		&5$\quad$$\,$Compute ${{P}_{l}^{(i)}}={{{\left| {{s}_{l}}\left( n \right) \right|}^{2}}}$\\
 		&6$\quad$ \textbf{if} $||\boldsymbol{P}^{i}-\boldsymbol{P}^{i-1}||_2\ge\varepsilon$\\
 		&7$\quad$ $\quad$ ${\boldsymbol{R}^{\left( i \right)}}=\boldsymbol{A}{\boldsymbol{P}^{\left( i \right)}}{\boldsymbol{A}^{H}}+\frac{1}{L}\sum\limits_{l=1}^{L}{\left| \frac{\boldsymbol{I}_{l}^{H}{\boldsymbol{R}^{\left( i-1 \right)}}^{\text{-}1}{{x}_{a}}\left( n \right)}{\boldsymbol{I}_{l}^{H}{\boldsymbol{R}^{\left( i-1 \right)}}^{\text{-}1}{\boldsymbol{I}_{l}}} \right|} ^{2}$\\
 		&8$\quad$ $\quad$ $i=i+1$\\
 		&9$\quad$ \textbf {else} $over$\\ 
 		&10$\,$$\,$$\,$\textbf{end}\\
 		&$\,$$\textbf{Output}$ $\boldsymbol{P}$\\
 		\bottomrule
 	\end{tabular}
 \end{table}
\section{SIMULATION RESULTS}
\label{sec:typestyle}
In this section, we use the computer to perform performance verification of the methods and all simulation results are based on the 2D virtual planar array in Fig.~\ref{fig2}, and $SNR= 20dB$. The FMCW simulation parameters table 1: 
\vspace{-0.5cm}
\begin{table}[h]
		\centering
		\caption{The simulation parameters}
	\label{tab11}
	\begin{tabular}{cclll}

			\toprule
		parameter & value  \\ \cline{1-2}
		RF carrier frequency $f_0$    &  79$GHz$ \\
		sweep bandwidth $B$        &      4$GHz$       \\
		number of fast time Sampling $N$     &    1024        \\  \cline{1-2}
	\end{tabular}
\end{table}

\begin{figure}[H]
	\centering
	\begin{minipage}[!]{0.5\linewidth }
		\centering
		\includegraphics[width= \linewidth,height=0.16\textheight]{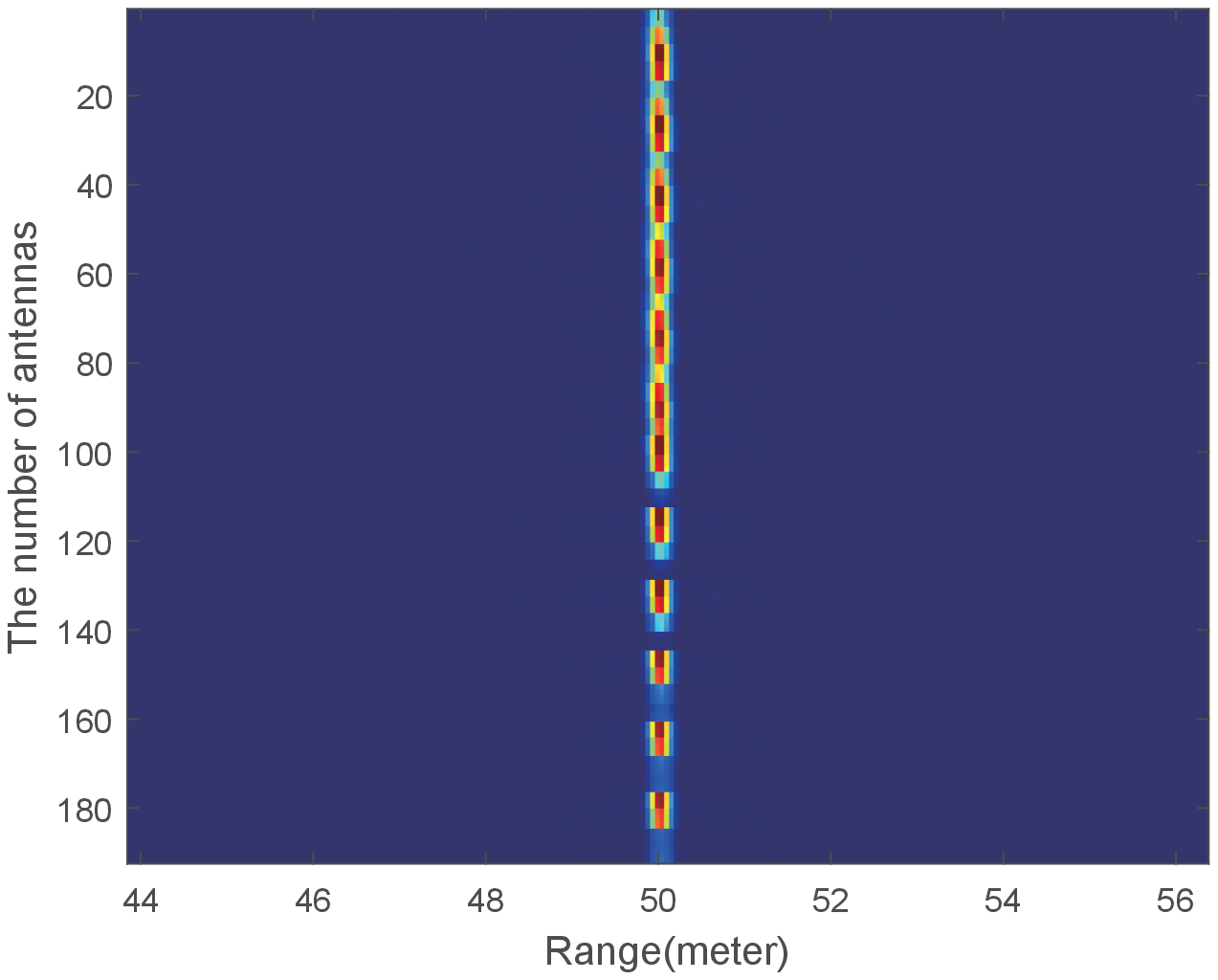}
		\begin{footnotesize}
			(a)
		\end{footnotesize}   
	\end{minipage}%
	\begin{minipage}[!]{0.5\linewidth }
		\centering
		\includegraphics[width= \linewidth,height=0.16\textheight]{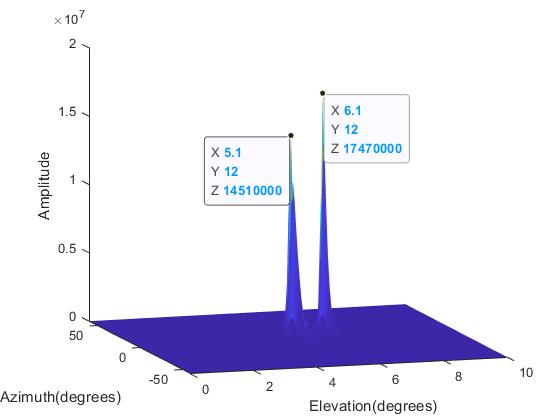}
		\begin{footnotesize}
			(b)
		\end{footnotesize}   
	\end{minipage}%
	\caption{ The results of two tragrts : (a) range imaging, (b) azimuth-elevation 2D super-resolution imaging.}
	\label{fig4}
\end{figure}

Due to limited paper, we mainly verify the azimuth super-resolution. Here, two point targets are simulated at different azimuth of  $\theta_1=5^{\circ }$, $\theta_2=6^{\circ }$, same elevation $\varphi=10^{\circ }$ in same range 50$m$. 
Through range FFT processing, we can get the target's range unit (50$m$) in Fig.~\ref{fig4} (a). And the 2D ADL-IAA algorithm is used to get the 2D super-resolution imaging with single-measurement in Fig.~\ref{fig4} (b). The azimuth angle is  $\theta ={{5.1}^{\circ }},{{6.1}^{\circ }}$ and elevation result is $\varphi ={{12}^{\circ }}$.

\begin{figure}[H]
	\centering
	\begin{minipage}[!]{0.5\linewidth }
		\centering
		\includegraphics[width= \linewidth,height=0.16\textheight]{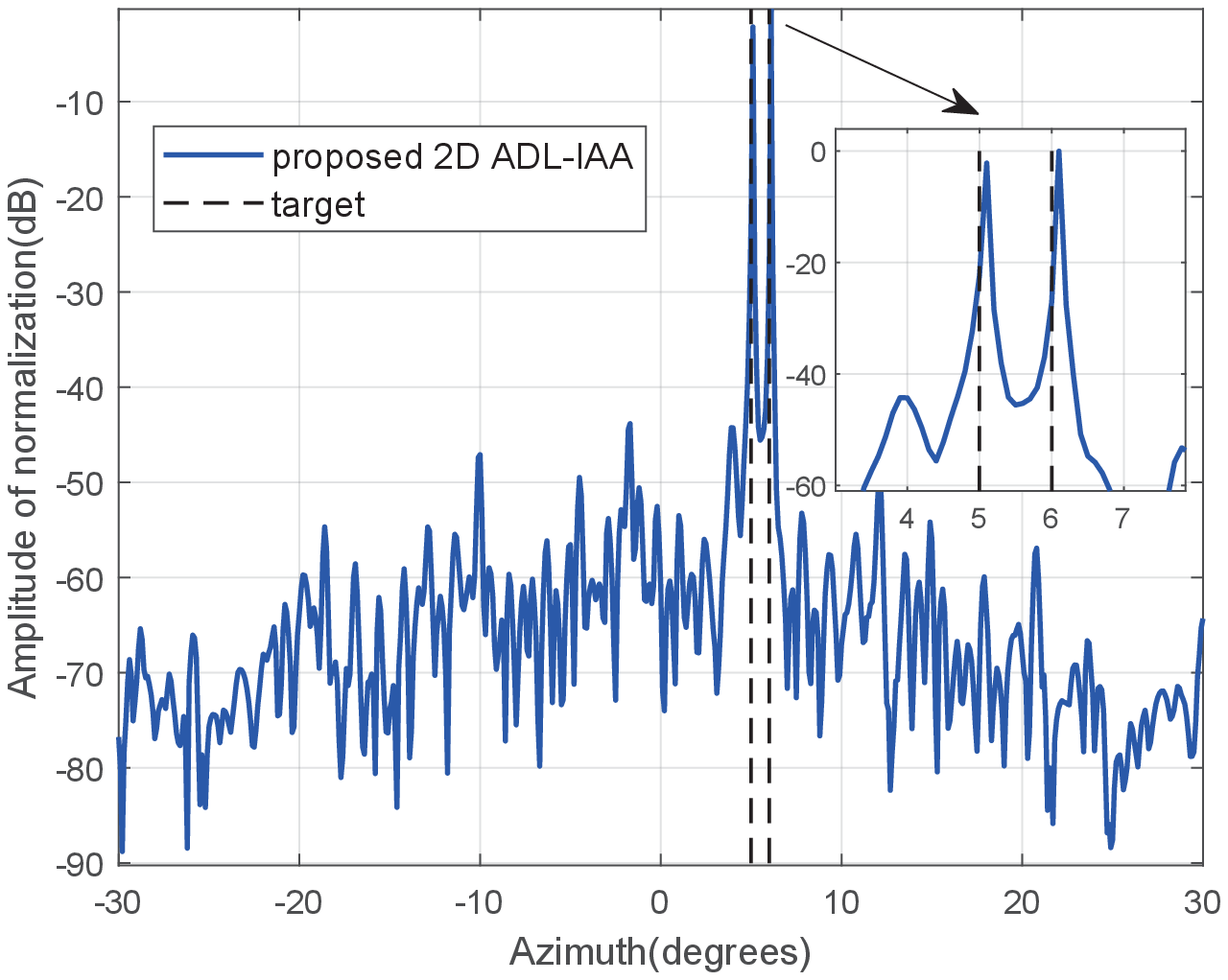}
		\begin{footnotesize}
			(a)
		\end{footnotesize}   
	\end{minipage}%
	\begin{minipage}[!]{0.5\linewidth }
		\centering
		\includegraphics[width= \linewidth,height=0.16\textheight]{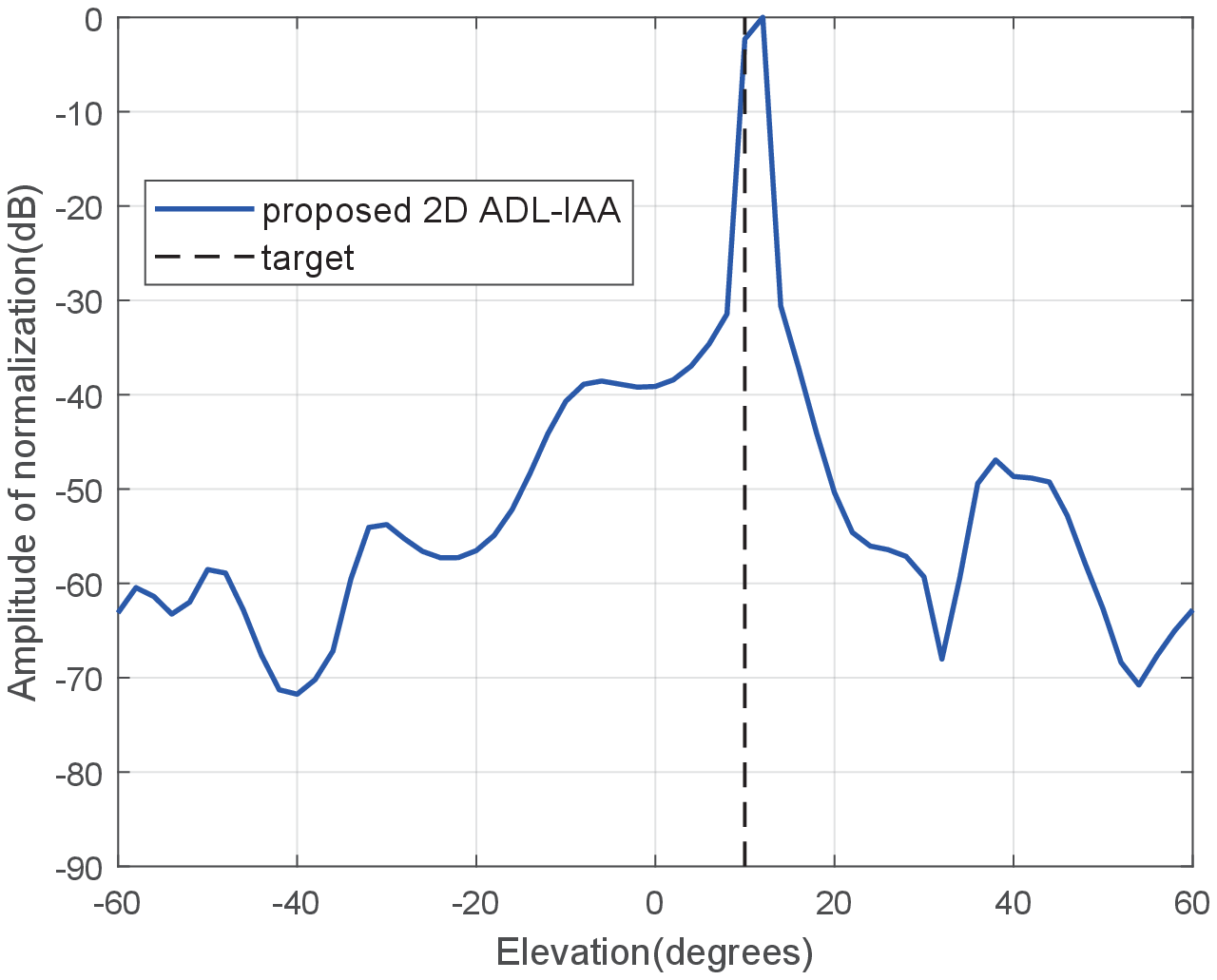}
		\begin{footnotesize}
			(b)
		\end{footnotesize}   
	\end{minipage}%
	\caption{ The azimuth and elevation profiles of 2D imaing : (a) azimuth result, (b) elevation result.}
	\label{fig5}
\end{figure}

Fig.~\ref{fig5} shows the azimuth and elevation profiles from azimuth-elevation of 2D imaging in Fig.~\ref{fig4} (b), and the proposed algorithm is performed in a single-measurement. From Fig.~\ref{fig5}(b), it is clear that the two targets with ${1}^{\circ }$ distance can be well separated using proposed 2D ADL-IAA. And it also has lower side lobes under -45$dB$, which is often beneficial to target perception by using CFAR processing. In Fig.~\ref{fig5} (b), the proposed 2D ADL-IAA can effectively estimate the elevation of the target with a small resolution error. Therefore, the simulation results proved the 2D ADL-IAA is able to work efficiently on this sparse array that is only a single-measurement, and it has super-resolution properties.
\begin{figure}[H]
	\centerline{\includegraphics[height=5cm]{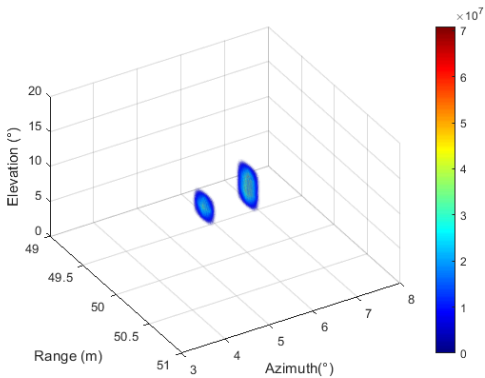}}
	\caption{The result of 3D super-resolution imaging.}
	\label{fig6}
\end{figure}
Finally, each range unit of range imaging is processed by the proposed 2D ADL-IAA. And after an appropriate threshold for CFAR is processing, using false alarm  ${10}^{-4}$. So, the final 3D imaging in Fig.~\ref{fig6} has the ability to score super-resolution through the distributed radar system.

\section{Conclusion}
\label{sec:majhead}
In this paper, we designed a new distributed mmW automotive radar system to improve imaging resolution. Then, we proposed a 3D super-resolution imaging method specifically for our system under the sparse and single-measurement conditions. Finally, the computer simulation results show the effectiveness of the method.

\end{document}